\documentstyle[aaspp4,12pt]{article}  
\begin{document}
%\renewcommand{\baselinestretch}{1.12}
%%%% begin new commands
%%%%%%%%%%%%%%%%%%%%%%%%%%%%% MACRO FILE %%%%%%%%%%%%%%%%%%%%%%%%%%%%%%%
%at the start of a tex file. Note that different printers may or
%may not automatically include an offset---if the text is off to
%one side adjust the commands \hoffset and \voffset by adding or
%removing a comment sign %.
\def\etal{et al.\ }
\def\chaphead{}
\def\absmath{\textfont0=\eightrm \scriptfont0=\sixrm
              \textfont1=\eightmit \scriptfont1=\sixmit}
\def\absfont{\let\rm=\eightrm \let\it=\eightit \rm\absmath}
\def\regmath{\textfont0=\twelverm \scriptfont0=\tenrm
              \textfont1=\twelvemit \scriptfont1=\tenmit}
\def\peterfont{\let\rm=\twelverm \let\it=\twelveit \rm\regmath}
%following macro defines vectors using arrows
\def\b#1{\skew{-2}\vec#1}  %This contradicts tex82 which has \b meaning
%underline instead of vector superscript. Skew is added to
%improve location of arrow.
%following macro defines vectors as boldface italic. Note in this case that
%all characters regarded as being in the argument of the macro will be
%boldface.

\def\emp{\sl}
\def\deffn{\bf}
\def\deg{^\circ}
\def\Vlasov{collisionless Boltzmann\ }
\def\lsls{\ll}
\def\grgr{\gg}
\def\erf{\mathop{\rm erf}\nolimits} %error function
\def\eqv{\equiv}
\def\real{\Re e}
\def\imag{\Im m}
\def\ctrline#1{\centerline{#1}}
\def\spose#1{\hbox to 0pt{#1\hss}}
\def\s{\ifmmode \widetilde \else \~\fi} %produces tilde in mathmode or
%horizontal mode.
     
\newcount\notenumber
\notenumber=1
\newcount\eqnumber
\eqnumber=1
\newcount\fignumber
\fignumber=1
     
\def\yyskip{\penalty-100\vskip6pt plus6pt minus4pt}
     
%\numberpara produces numbered paragraphs with extra space and no indentation
\def\numberpara{\yyskip\noindent}
     
\def\km{{\rm\,km}}
\def\kms{{\rm\,km\,s^{-1}}}
\def\kpc{{\rm\,kpc}}
\def\mpc{{\rm\,Mpc}}
\def\msun{{\rm\,M_\odot}}
\def\lsun{{\rm\,L_\odot}}
\def\rsun{{\rm\,R_\odot}}
\def\pc{{\rm\,pc}}
\def\cm{{\rm\,cm}}
\def\yr{{\rm\,yr}}
\def\au{{\rm\,AU}}
\def\AU{{\rm\,AU}}
\def\gm{{\rm\,g}}
\def\s{{\rm\,s}}
\def\dyne{{\rm\,dyne}}
\def\G{{\rm\,G}}
\def\erg{{\rm\,erg}}
\def\K{{\rm\, K}}
\def\rearth{{\rm\,R_\oplus}}
\def\mearth{{\rm\,M_\oplus}}

%\note macro produces sequentially numbered footnotes at bottom of page
%\foot macro produces sequentially numbered footnotes inserted in text
\def\note#1{\footnote{$^{\the\notenumber}$}{#1}\global\advance\notenumber by 1}
\def\foot#1{\raise3pt\hbox{\eightrm \the\notenumber}
     \hfil\par\vskip3pt\hrule\vskip6pt
     \noindent\raise3pt\hbox{\eightrm \the\notenumber}
     #1\par\vskip6pt\hrule\vskip3pt\noindent\global\advance\notenumber by 1}
\def\propo{\propto}
\def\larrow{\leftarrow}
\def\rarrow{\rightarrow}
          
%\Dt and \dt put Newton's notation dots above upper and lower case chars
\def\Dt{\spose{\raise 1.5ex\hbox{\hskip3pt$\mathchar"201$}}}    % upper case
\def\dt{\spose{\raise 1.0ex\hbox{\hskip2pt$\mathchar"201$}}}    % lower case
\def\llangle{\langle\langle}
\def\rrangle{\rangle\rangle}
\def\ldotss{\ldots}
\def\del{\b\nabla}
     
%reference macros. To generate reference to a paper in Ap.J. volume 300, p.123
%write \apj{Claus, S. 1990.}{300}{123}
\def\refindent{\par\noindent\parskip=4pt\hangindent=3pc\hangafter=1 }
\def\apj#1#2#3{\refindent#1.  {\sl Astrophys.\  J. }{\bf#2}, #3.}
\def\apjlett#1#2#3{\refindent#1.  {\sl Astrophys.\  J. Lett. }{\bf#2}, #3.}
\def\mn#1#2#3{\refindent#1.  {\sl Mon. Not. Roy. Astron. Soc. }{\bf#2}, #3.}
\def\mnras#1#2#3{\refindent#1.  {\sl Mon. Not. Roy. Astron. Soc. }{\bf#2}, #3.}
\def\aj#1#2#3{\refindent#1.  {\sl Astron. J. }{\bf#2}, #3.}
\def\aa#1#2#3{\refindent#1.  {\sl Astron. Astrophys. }{\bf#2}, #3.}
\def\Nature#1#2#3{\refindent#1.  {\sl Nature }{\bf#2}, #3.}
\def\Icarus#1#2#3{\refindent#1.  {\sl Icarus }{\bf#2}, #3.}
\def\refpaper#1#2#3#4{\refindent#1.  {\sl #2 }{\bf#3}, #4.}
\def\refbook#1{\refindent#1}
\def\science#1#2#3{\refindent#1. {\sl Science }{\bf#2}, #3.}
          
%\lta and \gta produce > and < signs with twiddle underneath
\def\lta{\mathrel{\spose{\lower 3pt\hbox{$\mathchar"218$}}
     \raise 2.0pt\hbox{$\mathchar"13C$}}}
\def\gta{\mathrel{\spose{\lower 3pt\hbox{$\mathchar"218$}}
     \raise 2.0pt\hbox{$\mathchar"13E$}}}
     
%\sec produces arcsec symbol so that 3\sec5 produces 3."5 with the second
%symbol and the period aligned.
\def\sec{\hbox{"\hskip-3pt .}}

%As for vectors: macro.tex is set up to show vectors with arrows on top. One
%writes $\b v$ to get a velocity vector. \b puts an arrow over the next single
%character. For the book, we are using Roman boldface to denote vectors. This
%uses the set of commands below in a macro file. These redefine \b so that now
%${\b v}$ produces boldface v. Note two warnings: 1. The brackets are
%essential, or else the rest of the formula will also be boldface. 2. The
%method doesn't work with lower case Greek letters, like \xi, \sigma, etc. For
%Wthese we have a new set of commands defined, \bxi, \bsigma, \btheta, \bomega,
%etc. Also \bnabla for bold face nabla, as in grad---as opposed to \nabla^2 for
%the Laplacian.

\font\syvec=cmbsy10                        %for boldface nabla
\font\gkvec=cmmib10                         %for boldface lowercase Greek

\def\bnabla{\hbox{{\syvec\char114}}}       %bold face nabla
\def\balpha{\hbox{{\gkvec\char11}}}        %bold face alpha
\def\bbeta{\hbox{{\gkvec\char12}}}         %bold face beta
\def\bgamma{\hbox{{\gkvec\char13}}}        %bold face gamma
\def\bdelta{\hbox{{\gkvec\char14}}}        %bold face delta
\def\bepsilon{\hbox{{\gkvec\char15}}}      %bold face epsilon
\def\bzeta{\hbox{{\gkvec\char16}}}         %bold face zeta
\def\boldeta{\hbox{{\gkvec\char17}}}       %bold face eta
\def\btheta{\hbox{{\gkvec\char18}}}        %bold face theta
\def\biota{\hbox{{\gkvec\char19}}}         %bold face iota
\def\bkappa{\hbox{{\gkvec\char20}}}        %bold face kappa
\def\blambda{\hbox{{\gkvec\char21}}}       %bold face lambda
\def\bmu{\hbox{{\gkvec\char22}}}           %bold face mu
\def\bnu{\hbox{{\gkvec\char23}}}           %bold face nu
\def\bxi{\hbox{{\gkvec\char24}}}           %bold face xi
\def\bpi{\hbox{{\gkvec\char25}}}           %bold face pi
\def\brho{\hbox{{\gkvec\char26}}}          %bold face rho
\def\bsigma{\hbox{{\gkvec\char27}}}        %bold face sigma
\def\btau{\hbox{{\gkvec\char28}}}          %bold face tau
\def\bupsilon{\hbox{{\gkvec\char29}}}      %bold face upsilon
\def\bphi{\hbox{{\gkvec\char30}}}          %bold face phi
\def\bchi{\hbox{{\gkvec\char31}}}          %bold face chi
\def\bpsi{\hbox{{\gkvec\char32}}}          %bold face psi
\def\bomega{\hbox{{\gkvec\char33}}}        %bold face omega

%to get bold face upper case greek characters, type {\bf \...}
% for example {\bf \Gamma} gives bold face upper case gamma.     
%%%%%%%%%%%%%%%%%%%%%%%%%%% END OF MACRO FILE %%%%%%%%%%%%%%%%%%%%%%%%%%
 
\def\ni{\noindent}
\def\L{{\cal L}}
%\lta and \gta produce > and < signs with twiddle underneath
\def\spose#1{\hbox to 0pt{#1\hss}}
\def\lta{\mathrel{\spose{\lower 3pt\hbox{$\mathchar"218$}}
     \raise 2.0pt\hbox{$\mathchar"13C$}}}
\def\gta{\mathrel{\spose{\lower 3pt\hbox{$\mathchar"218$}}
     \raise 2.0pt\hbox{$\mathchar"13E$}}}
\newcommand{\beq}{\begin{equation}}
\newcommand{\eeq}{\end{equation}}
\newcommand{\lp}{ \left(}
\newcommand{\rp}{ \right)}
\newcommand{\lc}{ \left[}
\newcommand{\rc}{ \right]}
\newcommand{\diff}{{\rm d}}
%%%% end new commands

\bibliographystyle{plain}

\title{Angular momentum redistribution by waves in the Sun}
\author{Pawan Kumar$^{1,2}$, Suzanne Talon$^{3,4}$ and Jean-Paul Zahn$^5$}
%\affil{Institute for Advanced Study, Princeton, NJ 08540}

\bigskip
\begin{abstract}

We calculate the angular momentum transport by gravito-inertial-Alfv\'en waves
and show that, so long as prograde and retrograde gravity waves are excited to
roughly the same amplitude, the sign of angular momentum deposit in the 
radiative interior of the Sun is such as to lead to an exponential growth
of any existing small radial 
gradient of rotation velocity just below the convection zone. 
This leads to formation of a strong thin shear layer (of thickness about 0.3\% 
R$_\odot$) near the top of the radiative zone of the Sun on a time-scale
of order 20 years. When the magnitude of differential rotation across 
this layer reaches about 0.1 $\mu$Hz, the layer becomes unstable to shear 
instability and undergoes mixing, and the excess angular momentum deposited 
in the layer is returned to the convection zone.
The strong shear in this layer generates toroidal magnetic field which is 
also deposited in the convection zone when the layer becomes unstable.
This could possibly start a new magnetic activity cycle seen at the surface.

\end{abstract}


\vskip 2.5truecm
$^1$Institute for Advanced Study, Princeton, NJ 08540

$^2$A.P. Sloan fellow

$^3$D\'epartement de Physique, Universit\'e de Montr\'eal, Montr\'eal PQ H3C 3J7

$^4$CERCA, 5160 boul. D\'ecarie, Montr\'eal PQ H3X 2H9

$^5$Observatoire de Paris, Section de Meudon, 92195 Meudon, France
\vfill\eject

\section{Introduction} 
\baselineskip=15pt

The Sun is currently losing angular momentum at its surface via a wind
at a rate of the order of $10^{31}$ g cm$^2$ s$^{-2}$, which 
is slowing down the surface layers. However, we know from
the frequency splitting of p-modes that the radiative interior of the
Sun is rotating as a solid body at a rate that is not very different
from the mean surface rotation rate. Thus the loss of angular momentum at
the surface is communicated to the rest of the Sun on a time-scale of
the order of the solar age or less.

The convective motions in the outer third of the Sun can very efficiently
redistribute angular momentum on a short convective time-scale of about
a month. However, processes responsible for the redistribution of angular 
momentum in the radiative interior are less certain. One possibility is that 
magnetic torques can extract angular momentum from the radiative interior.
Charbonneau \& MacGregor (1993) have carried out a detailed numerical analysis 
of this transport process. However, there are some drawbacks to that 
mechanism, which are discussed in Zahn (1997).

Another possibility is that the gravity waves generated in the convection
zone can extract/deposit angular momentum in the radiative interior.
The angular momentum flux carried by these waves is enormous and they 
are very efficient in redistributing angular momentum in the 
radiative interior (Schatzman 1993a-b; Kumar \& Quataert 1997; 
Zahn, Talon \& Matias 1997).

Here, we explore this second mechanism in some detail. 
In particular, we discuss the effects of the
Coriolis force and of magnetic fields on the dispersion relation of gravity 
waves and on the deposition of angular momentum in the radiative
interior on the Sun (\S 2). We describe how these waves lead to the 
formation of a strong and thin shear layer just below the convection zone 
which becomes unstable when the gradient exceeds a critical value leading 
to mixing of elements and angular momentum with the
convection zone (\S 3). We further discuss how this shear layer
could contribute to the generation of magnetic field and perhaps
to the magnetic cycle observed in the Sun (\S 4).

\section{Angular momentum transport by waves}

We derive below a general expression for the angular momentum flux in 
waves. The dispersion relation for gravito-inertial waves in the presence
of a magnetic field is calculated in \S 2.2.

\subsection{Angular momentum flux in waves}

Let us consider a fluid Lagrangian density ${\cal L}$ which is a function of
the displacement field $\bxi$, and of its temporal and spatial derivatives.
If ${\cal L}$ is not an explicit function of the azimuthal coordinate $\phi$,
the z-component of the angular momentum is conserved (this is a particular
case of the Noether's theorem cf. Quigg, 1983). The
Lagrangian density thus obeys the relation
\beq
   {d\L\over d\phi} = 0 = {\partial \L\over \partial\phi} + {\partial \L\over
   \partial\xi_i} {\partial\xi_i\over\partial\phi} + {\partial \L\over
   \partial\xi_{i,t}} {\partial^2\xi_i\over\partial\phi\partial t} +
   {\partial \L\over \partial\xi_{i,j}} {\partial^2\xi_i\over\partial
   \phi\partial x_j},
\eeq
or
\beq
   {d\L\over d\phi} = {\partial\over\partial t}\left[{\partial \L\over\partial
   \xi_{i,t}} {\partial\xi_i\over\partial\phi}\right] + {\partial\over
   \partial x_j} \left[{\partial \L\over\partial\xi_{i,j}}
   {\partial\xi_i\over\partial\phi}\right]
  - {\partial\xi_i\over\partial\phi} \left[{\partial\over \partial t}
    {\partial \L\over\partial\xi_{i,t}} + {\partial\over \partial x_j}
    {\partial \L\over\partial\xi_{i,j}} - {\partial \L\over\partial
   \xi_i}\right] = 0.
\eeq
It follows from the Euler-Lagrange equation that the last term is zero
leading to
\beq
   -{\partial\over\partial t}\left[{\partial \L\over\partial\xi_{i,t}} {\partial
   \xi_i\over\partial\phi}\right] + {\partial\over\partial x_j} \left[
   g_{\phi j} \L - {\partial \L\over\partial\xi_{i,j}} {\partial\xi_i
   \over\partial\phi} \right] = 0.
\eeq
This equation expresses the conservation of wave angular momentum. The 
angular momentum density and flux are given by:
\beq
   {\cal L}_{am} = - {\partial \L\over\partial\xi_{i,t}}{\partial\xi_i
   \over\partial\phi}, \label{lam}
\eeq
and
\beq
  {\cal F}^{(i)}_{am} = g_{\phi j} \L - {\partial \L\over\partial\xi_{i,j}}
  {\partial\xi_i\over\partial\phi}. \label{fam}
\eeq
We show below that the above expression for angular momentum density 
(${\cal L}_{am}$) has the correct sign and magnitude.

Consider a plane wave in a homogeneous medium traveling along +x direction,
\beq
\bxi = \xi_0 \hat x \sin(\omega t - k x).
\eeq
The only term in the Lagrangian containing
$\partial\bxi/\partial t$ is the kinetic energy term, so it
is straightforward to calculate the angular momentum density
for the wave using eq.~(\ref{lam}):
\beq
  {\cal L}_{am} = - \rho {\partial\bxi\over\partial t}\cdot
  {\partial\bxi\over\partial\phi} = - \rho\xi_0^2\omega k r\sin\phi\cos^2
   (\omega t - k x)=- \rho \left({k\over\omega}\right) \left({\partial
   \bxi\over\partial t}\right)^2 y, \label{lhom}
\eeq
where $y=r\sin\phi$.
Since the wave is traveling along the $+x$ direction, its linear momentum
density is also along the $+x$ direction and has the magnitude
$p = \rho (\omega\xi_0)^2 (k/\omega)$.
The angular momentum density of the wave should be ${\cal L}_{am} = 
{\bf x}\times {\bf p} = -\rho u^2 (k/\omega) y\hat z$, which is identical 
to the expression~(\ref{lhom}).

The wave displacement field in the spherical Sun can be written as
\beq
  \bxi=\left( \xi_r, \xi_h{\partial\over\partial\theta}, {\xi_h\over\sin\theta}
   {\partial\over\partial\phi}\right) Y_{\ell m}(\theta, 0)\cos(\omega t -
 m\phi).
\eeq
Substituting this in the expression for ${\cal L}_{am}$ (eq.~\ref{lam}),
we obtain
\beq
   {\cal L}_{am} = \rho m\omega\left\{\left[ \xi_r^2Y_{\ell m}^2 + \xi_h^2
   \left({\partial Y_{\ell m}\over\partial\theta}\right)^2\right]
   \sin^2(\omega t - m\phi) + {m^2\xi_h^2\over\sin^2\theta} Y_{\ell m}^2
   \cos^2(\omega t - m\phi)\right\}.
\eeq
Integrating ${\cal L}_{am}$ over $\theta$ and $\phi$ we obtain
\beq
   \langle{\cal L}_{am}\rangle = {\rho m\omega\over 2}\left[ \xi_r^2 +
   \ell(\ell+1)\xi_h^2\right] = {m{\cal E}\over\omega},
\eeq
where ${\cal E}$ is the energy density in the wave.
Prograde waves ($m>0$) thus carry {\em positive} angular momentum
whereas retrograde waves ($m<0$) carry {\em negative} angular 
momentum\footnote{The sign of the angular momentum carried by the
gravity waves was improperly defined in  Kumar \& Quataert (1997)
and in Zahn, Talon \& Matias (1997), as was pointed out by Ringot (1998)}. 

Waves deposit their angular momentum in the star only when and where 
they are damped. For gravity waves, the
main damping process is the photon diffusion, which is discussed in
\S2.4. The angular momentum flux crossing a spherical shell may then 
be expressed as a function of depth:
\beq
{\cal L}_{am} (\omega, \ell, r) = {\cal L}_{am}(\omega, \ell, r_{c})\,  
\exp[ -\tau(\omega, \ell, r)], \label{Lam}
\eeq
where $\tau(\omega, \ell, r)$ is the damping ``optical depth'' for a wave of
frequency $\omega$, and degree $\ell$, between the radius $r$ and $r_{c}$
(the radius of the bottom of the convection zone) cf. \S2.4.

\subsection{Dispersion relation for Gravito-Inertial-Alfv\'en waves}

The linearized momentum equation that includes both
Coriolis and Lorentz forces is
\beq
 \rho {\partial^2{\bxi} \over \partial t^2} + {\bnabla} p_1 - \rho_1
{\bf g} + 2 \rho {\bf \Omega} \times {\bf v} = {\left({\bnabla}
\times {\bf B}_1\right)\over 4\pi}\times {\bf B}
\eeq
where $\bf \Omega$ is the rotational frequency, ${\bf g} = -g \hat{\bf r}$
is the gravitational acceleration, $\bxi$ and ${\bf v}$ are the
fluid displacement and velocity associated with the wave, 
and $\rho_1$, $p_1$ and ${\bf B}_1$ are the Eulerian perturbations
of density, pressure and magnetic field associated with the wave. 
We assume for simplicity that the unperturbed magnetic field ${\bf B}$ and
the rotational speed ${\bf \Omega}$ are constant.
In the short wavelength limit, we can take the spatial and the temporal 
dependence of all of the perturbed quantities to be proportional to 
$\exp(i{\bf k}\cdot{\bf x} - i\omega t)$. Substituting this in the above
equation we obtain
\beq
-\rho\omega^2{\bxi} + i{\bf k} p_1 - \rho_1 {\bf g} -
2 i\omega\rho\, {\bf \Omega} \times {\bxi} = {i\left({\bf k} \times
{\bf B}_1\right)\over 4\pi}\times {\bf B}.
\eeq
Making use of the linearized equation of entropy conservation
\beq
\rho_1 = {p_1\over c^2} + {\rho\over g} N^2\xi_r\ , \label{lee}
\eeq
where $\xi_r$ is the radial component of the displacement vector, 
and taking $k\gg g/c^2$, i.e. the wavelength is much smaller than 
the density scale height, we obtain
\beq
-\rho\omega^2{\bxi} + i{\bf k}p_1 + \rho
N^2\xi_r\hat{\bf r} - 2 i\omega\rho{\bf \Omega} \times {\bxi} =
{i\left({\bf k}\times{\bf B}_1\right)\over 4\pi} \times
{\bf B}. \label{xir}
\eeq
Substituting for ${\bf B}_1$ from the linearized induction equation for a 
perfectly conducting fluid, and making use of the incompressibility condition, 
${\bf k} \cdot{\bxi} = 0$, we obtain
\beq
-\rho\omega^2{\bxi} + i{\bf k} p_1 + \rho N^2\xi_r\hat{\bf r} - 2
i\omega\rho\left({\bf \Omega} \times {\bxi}\right) + {\left({\bf
k} \cdot{\bf B}\right)\over 4\pi} \left\{\left({\bf k} \cdot
{\bf B}\right){\bxi} - \left({\bxi} \cdot {\bf B}\right){\bf k}\right\} = 0.
\eeq
The vector cross product of the above equation with {\bf k} yields
\beq
\left(\omega^2 - ({\bf k}\cdot \hat{\bf B})^2 V^2_A\right)\left({\bf k} \times
{\bxi}\right) - N^2\xi_r\left({\bf k} \times \hat{\bf r}\right)- 2
i\omega\left({\bf k} \cdot{\bf \Omega}\right){\bxi} = 0, \label{vcp}
\eeq
where $V^2_A = {B^2/ 4\pi\rho}$ is the Alfv\'en velocity, and $\hat{\bf B} = 
{{\bf B}/B}$.
The vector cross product of eq. (\ref{vcp}) with {\bf k} gives
${\bf k}\times\bxi$ in terms of $\bxi$, which, when substituted back
into eq. (\ref{vcp}), results in
\beq
 {\bxi}\left[{\left(\omega^2 - ({\bf k}\cdot \hat{\bf B})^2 V^2_A\right)^2k^2
 \over 2 i\omega\left({\bf k} \cdot {\bf\Omega}\right)} + 
2i\omega\left({\bf k} \cdot {\bf \Omega}\right)\right]
 + \xi_r\left[N^2\left({\bf k} \times \hat{\bf r}\right) +
{N^2\left(\omega^2 - ({\bf k}\cdot \hat{\bf B})^2 V^2_A\right)\over 2 i 
\omega \left({\bf k} \cdot{\bf\Omega}\right)} {\bf k} \times \left({\bf k}
\times \hat{\bf r}\right)\right] = 0.
\eeq
The $r$-component of the above equation yields the desired 
dispersion relation:
\beq
\left(\omega^2 - ({\bf k}\cdot \hat{\bf B})^2 V^2_A\right)^2 -
\left({\bf N} \times \hat{\bf k}\right)^2\left(\omega^2 -
({\bf k}\cdot \hat{\bf B})^2 V^2_A\right) - 4\omega^2\left(\hat{\bf k} \cdot{\bf
\Omega}\right)^2 = 0, \label{dispers}
\eeq
where $\hat{\bf k} = {{\bf k}/k}$.
This equation can be solved easily to determine the wave frequency
\beq
\omega^2 = ({\bf k}\cdot \hat{\bf B})^2 V^2_A + {1\over 2} \Biggl[ 
({\bf N} \times \hat{\bf k})^2 + 4(\hat{\bf k} \cdot{\bf \Omega})^2 \pm
\left\{ \left[ ({\bf N} \times \hat{\bf k})^2 + 4(\hat{\bf k} \cdot{\bf \Omega})^2
  \right]^2 + 16 ({\bf k}\cdot \hat{\bf B})^2 V^2_A (\hat{\bf k} \cdot
{\bf \Omega})^2\right\}^{1/2}\Biggr].
\eeq

This dispersion relation has two branches. The lower frequency branch
corresponds to Alfv\'en waves and the high frequency branch, to 
gravito-inertial-Alfv\'en waves. The dispersion relation for the high 
frequency branch can be written as
\beq
   \omega^2 \approx ({\bf k}\cdot \hat{\bf B})^2 V^2_A  + ({\bf N} \times 
   \hat{\bf k})^2 + 4(\hat{\bf k} \cdot{\bf \Omega})^2. 
    \label{disprel}
\eeq

The radial wave number $k_r$, when the magnetic field lies in the
horizontal plane, can be obtained from the dispersion relation 
(eq. \ref{dispers}) and is given by
\beq k_r = {-2\omega^2\Omega^2 k_\theta\sin2\theta + \left[4\omega^4\Omega^4 
  k_\theta^2 \sin^22\theta + (\omega_1^4 - 4\omega^2\Omega^2\cos^2\theta)
  \left\{(N^2-\omega_1^2) \omega_1^2 k_h^2 + 4\omega^2\Omega^2 
  k_\theta^2\sin^2\theta\right\}\right]^{1/2} \over \omega_1^4 
  - 4\omega^2\Omega^2\cos^2\theta}, \label{kr}
\eeq
where 
\beq
 \omega_1^2 = \omega^2 - ({\bf k}\cdot \hat{\bf B})^2 V^2_A,
\eeq
${\bf \Omega} = \Omega(\cos\theta\, \hat{\bf r} - \sin\theta\;\hat{\btheta})$,
$k_h$ is the horizontal wavenumber, $k_\theta^2 = k_h^2 - k_\phi^2$,
$k_\phi = m/r\sin\theta$, and $m$ is the azimuthal number.
Figure 1 shows $k_r$ as a function of $\omega$ for a few cases of
$\theta$, $B$ and $k_h$.

The frequency $\omega$ in the expression for the wavenumber is the frequency
as seen in the local rest frame of the fluid i.e. $\omega = \omega_I - 
m\Omega$, where $\omega_I$ is the wave frequency in an inertial frame.
The component of the wave-vector along the $\phi$ direction being
conserved along the ray path, the prograde waves moving inwards in a 
region of increasing rotation speed are Doppler shifted to lower frequencies.
 From the above equation, we see that the wavenumber of these waves increases
with decreasing frequency (see also fig. 1) therefore enhancing their damping.
Thus, in a region of increasing rotation speed, prograde waves are 
dissipated more strongly than retrograde waves. 

%Suz - corrected
For pure gravity waves, Zahn et al. (1997) discussed the existence of 
a critical layer determined by the condition $\omega = m\delta\Omega$,
where $\delta\Omega$ is the difference between the rotation rate at
radius $r$ and the base of the convection zone. 
In that case, a wave is completely absorbed since its radial wave 
number diverges and its group speed goes to zero. The presence of 
non-zero magnetic field and/or rotation modifies the condition for 
the appearance of critical layers, which can occur even when 
$\omega>0$ in the local rest frame. However, it can be shown that 
$\omega_1^2 = \pm 2\omega\Omega\cos\theta$, for which the denominator 
in equation (\ref{kr}) is zero, does not correspond to a critical layer 
since the numerator at this frequency is zero as well, unless either 
$\Omega\cos\theta=0$ or the wave frequency in the local rest 
frame of the fluid ($\omega - m\delta\Omega$) is zero.
\footnote{Barnes et al. (1998) discuss critical layers for 
gravito-Alfv\'en waves and their results can be obtained 
from ours by setting $\Omega=0$.}
For the low frequency waves (which carry most of the angular momentum)
radiative damping is very strong and the waves are damped on a
distance scale much smaller than the solar radius even when waves do
not encounter a critical layer. This is discussed in more detail in \S2.4.

Wave propagation requires $\omega> ({\bf k}\cdot \hat{\bf B}) V_A$. 
For horizontal magnetic field of strength less than about
(2$\cdot$10$^5/\ell)$ Gauss wave propagation in the radial direction
requires the frequency to be greater than 0.5 $\mu$Hz,
while for a radial magnetic field 
of strength less than about 10$^3$ Gauss the condition for wave propagation 
is $\omega> 0.5 \mu$Hz (angular momentum flux in waves peaks at 
the convective frequency of 0.3 $\mu$Hz and falls off as $\nu^{-5.5}$
at higher frequencies). Waves will undergo reflection when 
the radial component of the magnetic field is sufficiently large.

\bigskip
\subsection{Excitation of waves by turbulent convection}
\medskip

Gravity waves with periods of about 10 days are expected to be excited 
in the Sun by a number of different processes such as the Reynold's 
stress in the convection zone, plumes penetrating in the stably stratified
radiative interior, etc. In the Earth's atmosphere, the latter process
is known to be the most efficient and it might be the dominant process
in the Sun as well. However, the energy flux in gravity waves resulting
from this process is subject to great uncertainty because
we know little about the properties of plumes at the base of the solar
convection zone. We will therefore consider wave excitation only by 
Reynold's stress. The results of this paper (the
formation of shear layer at the top of the radiative interior
of the Sun by gravity waves) can be easily 
modified to incorporate more efficient gravity 
waves generation by plumes or some other process. Generally, larger flux in 
gravity waves means more efficient formation of shear layer
on a smaller time-scale.

The energy flux per unit frequency in waves just below the convective 
envelope of the Sun, as a result of excitation by the Reynold's
stress, can be calculated using the following expression (cf. Goldreich
et al. 1994):
\beq
F_E^{(c)} = {\omega^2\over 4\pi} \int dr\; {\rho^2\over r^2} 
   \left[\left(\frac{\partial \xi_r}{\partial r}\right)^2 + 
   \ell(\ell+1)\left(\frac{\partial \xi_h}{\partial r}\right)^2 \right] 
   \exp\left[ -h_\omega^2 \ell(\ell+1)/2r^2\right] \frac{v^3 L^4 }{1 
  + (\omega \tau_L)^{15/2}},\label{eflux}
\eeq
where 
$\xi_r$ and $[\ell(\ell+1)]^{1/2}\xi_h$ are the radial and horizontal
displacement wavefunctions which are normalized to unit energy flux just 
below the convection zone, $v$ is the convective velocity, $L$ is the radial
size of an energy bearing turbulent eddy, $\tau_L \approx L/v$ is the
characteristic convective time, and $h_\omega$ is the
radial size of the largest eddy at $r$ with characteristic frequency of
$\omega$ or greater ($h_\omega = L \min\{1, (2\omega\tau_L)^{-3/2}\}$).
The gravity waves are evanescent in the convection zone, the region
where they are excited.
The above equation was derived under the assumption that the
turbulence spectrum is Kolmogorov and, as mentioned earlier,
it ignores wave excitation resulting from convective overshooting.

The energy flux spectrum just below the convection zone is 
calculated numerically using equation (\ref{eflux}) and the result is
shown in fig. 2. Also shown in fig. 2 is the 
frequency integrated  wave action as a function of wave degree $\ell$;
angular momentum flux for waves of a given $m$ is equal to $m$ times the
wave action. We provide below a rough estimate of the energy flux in 
gravity waves in order to gain some understanding of the numerical results.

For the low frequency waves of interest here, the WKB solution to the
wave equation is quite accurate and can be used to calculate the energy
flux. The displacement wavefunctions $\xi_r$ and $\xi_h$ in the short
wavelength limit are proportional to $k_r^{-1/2}\exp(i\int dr k_r)$, 
and the energy flux in waves is $\rho\omega^2(\xi_r^2 + \ell(\ell+1)\xi_h^2) 
v_g$, where $k_r$ is radial wave number given by equation (\ref{kr}), 
and $v_g$ is the wave group velocity. 

Using the continuity of the radial velocity at the interface of the
radiation and the convection zone we find that the properly normalized
radial displacement wave function in the convection zone is
\beq 
\xi_r \sim {(k_h N_c)^{1/2} \over \omega (\rho N N_t)^{1/2} }
 \exp\Bigl[-\int dr\, k_r\Bigr],\label{xir} 
\eeq
where $N_t\sim (g/H)^{1/2}$ is the Brunt-V\"ais\"al\"a frequency at the top of 
the radiation zone and $N_c$ is the convective frequency at the bottom
of the convection zone. Substituting this into equation (\ref{eflux})
and making use of $L\sim H$ and $\rho v^3 \sim F_\odot$ (the energy
flux carried by convection), we find the energy flux in waves to be
\beq
F_E^{(c)}(\omega) \sim {F_\odot k_h^3 H_\omega^{5}\,\exp(-H_\omega^2 k_h^2)
 \over r_c^2 N_t}. 
\eeq
The subscript $\omega$ on $H$ emphasizes that it is the scale height at 
a location in the convection zone where the characteristic time-scale of 
energy bearing eddies is $\omega^{-1}$. For a polytropic atmosphere of index 
$n$, the density scales with depth $z$ (measured from the surface) as $z^n$, and 
therefore the convective velocity scales as $v \propto z^{-n/3}$.
Thus the convective frequency $\omega\propto z^{-(n+3)/3}$,
and the scale height $H_\omega \sim H_c (N_c/\omega)^{3/(n+3)}$; $H_c$
is the scale height at the bottom of the convection zone.
Substituting this into the above equation we obtain
\beq
F_E^{(c)}(\omega) \sim {F_\odot k_h^3 H_c^5\, \exp(-H_\omega^2 k_h^2)\over
 r_c^2 N_t} \left( {\omega\over N_c}\right)^{-15/(n+3)}.
\eeq
Since the polytropic index of the convection zone is approximately 1.5,
we find the frequency dependence of $F_E^{(c)}$ to be $\omega^{-3.3}$.
The numerical calculation of the energy energy flux, using equation
(\ref{eflux}), gives the frequency dependence to be
$\omega^{-4.5}$. The difference arises because of our neglect of
the frequency dependence of $k_r$ in equation (\ref{xir}) for $\xi_r$.

Integrating $F_E^{(c)}$ over the wave frequency and the 
horizontal wave number, we find the total energy flux in waves to be 
given by\footnote{Goldreich and Kumar (1990) had obtained the same
result for the energy flux in surface gravity waves.}
\beq
F_E^{(c)} \sim {F_\odot N_c\over N_t}\sim F_\odot {\cal M}_t,
\eeq
where ${\cal M}_t\approx 3\times 10^{-4}$ is the turbulent Mach number 
at the bottom of the convection zone. Thus the total energy flux in
the long period gravity waves just below the solar convection zone is
about 0.03\% of the solar luminosity; more accurate numerical calculation
based on the evaluation of eq. (24) gives the energy flux to be 0.1\% of 
the solar luminosity.

\bigskip
\subsection{Radiative damping of waves}
\medskip

For the long period or short wavelength waves which dominate the transport 
of angular momentum, the radiative diffusion of photons from regions of
positive and negative temperature fluctuation across a 
wavelength is the dominant damping process. The damping opacity 
between radii $r$ and $r_c$ (the base of the convection
zone) is calculated using:
\beq
\tau(\omega, \ell, m, r) = \int_r^{r_c} dr'\, {\gamma(\omega_*^{(m)}(r'),
   \ell, r')\over v_g(\omega_*^{(m)}(r'), \ell, r')}
\eeq
where
\beq 
\omega_*^{(m)}(r) = \omega + m\left[ \Omega_c - \Omega(r)\right],
\eeq
is the wave frequency in the local rest frame of the fluid at radius $r$,
$\Omega_c$ is the angular rotation speed at the base of the convection zone, 
the damping rate
\beq
\gamma(\omega, \ell, m, r) \approx K k_r^2
\eeq
\beq 
K = {16\sigma T^3\over 3 \rho^2\kappa c_p}\approx {2 F_r H_T\over 5 p},
\eeq
is the thermal diffusivity, $\sigma$ is the Stefan-Boltzmann constant,
$\kappa$ is opacity per unit mass, $c_p$ is the specific heat per unit mass,
$F_r$ is the radiative energy flux, $H_T$ is the temperature scale height,
$p$ is the gas pressure, and $k_r$ is the radial wavenumber of the wave.

For gravity waves, $k_r\approx k_h N/\omega$ and $v_g\approx \omega^2/(k_h N)$.
Therefore, their damping increases rapidly with decreasing 
frequency and increasing $\ell$ as $\omega^{-4}\ell^3$. The wave damping length
$d_w \sim v_g/\gamma \approx 5p\omega/(2F_r H_T k_r^3)$, and so the ratio of
the damping length and the wavelength is
\beq
d_w k_r \sim {\rho \omega^3 H_T\over F_c k_h^2} \sim \left({\omega\over
   N_c}\right)^3 \left({r_c\over H_T}\right)^2 \ell^{-2},
\eeq
where $N_c$ is the convective frequency at the base of the convection 
zone. We have used the relations $F_r\sim \rho (N_c H_T)^3$ and 
$N^2\sim g/H_T$ in deriving the above scaling. At the bottom of the solar 
convection zone $r_c/H_T\sim 10$, therefore for dipole waves of frequency
$N_c$ the damping length is about 15 wavelengths. For waves of
frequency 0.6 $\mu$Hz ($\omega/N_c\approx 4$) the wave damping length
is of order the wavelength for $\ell\sim 20$ (see fig. 3 for a more 
accurate numerical result).

Numerical calculation of damping length as a function of wave frequency 
and $\ell$ are shown in figure 3. Note that, even in the absence of a 
critical layer wave damping is extremely efficient (see also Press, 1981). 
Waves with frequency of 0.5$\mu$Hz and $\ell=15$ are damped on a length scale 
of about 10 wavelengths. Our numerical calculations for the evolution of
shear layer below the convection zone, presented in \S3, include radiative
damping of waves throughout the shear layer and not just at the critical
surface. 

Raditive damping only affects the gravity wave part of the 
gravito-Alfv\'en-inertial waves. Thus, after traversing some distance,
a gravito-Alfv\'en-inertial wave
can be converted into an Alfv\'en-inertial wave 
as a result of radiative damping. 
We have however ignored this complication here because, for
the waves of frequency greater than about 0.6 $\mu$Hz
considered in this paper, the buoyancy force is more important than
magnetic tension and the Coriolis forces as long as the toroidal magnetic
field below the convection zone is less than 10$^5$ Gauss, and the 
latitute is less than 45$^o$ (see \S2.2). Thus, much of the angular
momentum flux carried by gravito-Alfv\'en-inertial waves gets deposited
in the fluid before they turn into Alfv\'en-inertial waves. Moreover,
it can be shown that for parameters of interest to the solar problem, 
the finite plasma conductivity damps out Alfv\'en-inertial waves near the
critical surface.

\bigskip
\section{The formation and evolution of a shear layer}
\medskip

Long period gravity waves generated at the base of the solar convection zone 
are efficient at redistributing the angular momentum in the radiative 
interior (Zahn, Talon \& Matias, 1997, Kumar \& Quataert 1997).
The angular momentum deposited by gravity
waves just below the convection zone is such as to enhance any
pre-existing gradient of angular velocity. The physical reason for
this, although not intuitive, is easy to understand. Consider the 
case where the angular 
speed increases with depth below the convection zone. The frequencies
of prograde waves moving inward in the radiative interior are decreased as
seen in the local rest frame of the rotating fluid whereas the frequencies
of retrograde waves are increased. Thus the damping of prograde waves is
enhanced relative to retrograde waves and, as a result, positive angular
momentum is deposited just below the convection zone enhancing the magnitude 
of the existing differential rotation. This enhancement of angular momentum
gradient is confined to a thin layer whose thickness is of the order 
of the wave damping
length. Initially, the angular speed increases exponentially with time, 
and when the gradient becomes sufficiently large so that much of the 
angular momentum flux of prograde waves is absorbed in the shear layer 
below the convection zone, further growth becomes linear in time. 
Retrograde waves, which
were only partially absorbed in this layer, continue to propagate deeper
in the radiative interior and deposit their negative angular momentum 
in a thin layer which is spun down and develops a strong differential
rotation.

The evolution of angular momentum and formation of the shear layer are
determined by solving the following equation:
\beq 
\rho r^2{d\Omega\over dt} = {d F_L\over dr}, \label{domega}
\eeq
where $F_L$ is the angular momentum flux associated with waves
\beq
F_L = \sum_{\ell, m} \int d\omega\, {m F_E^{(c)}(\omega, \ell)\over\omega}
 \exp\left[-\tau(\omega, \ell, m)\right]. \label{FL}
\eeq

To estimate the time-scale for the initial exponential growth
of the angular speed, we expand the damping optical depth
$\tau$ 
\beq
\tau(\omega, \ell, m) = \int_r^{r_c} dr\, \gamma(r, \omega-m\delta\Omega, \ell)
 \approx \int_r^{r_c} dr\, \gamma(r, \omega, \ell) - \int_r^{r_c} dr\, 
  m\delta\Omega {d\gamma\over d\omega} \equiv \tau_0(r, \omega, \ell) +
   m \tau_1(r, \omega, \ell).
\eeq
Substituting this back into equation (\ref{FL}) we obtain
\beq 
F_L = -\sum_\ell \int d\omega {F_E^{(c)}\over\omega} {d\over d\tau_1}
  \left[{\sinh(\ell+1/2)\tau_1\over\sinh(\tau_1/2)}\right] \exp(-\tau_0)
 = \sum_\ell \int d\omega {\tau_1 F_E^{(c)}(\ell+1/2)^3\over 3\omega}
  \exp(-\tau_0).
\eeq
The last equality was obtained by assuming that $\ell\tau_1\ll 1$
and $\ell\gg 1$. Finally, substituting this in equation (\ref{domega}), 
we obtain
\beq 
{d\delta\Omega\over dt} \approx {\delta\Omega\over \rho r^2} \sum_\ell
  \int\, d\omega {d\gamma\over d\omega} \exp(-\tau_0) 
   {F_E^{(c)}(\ell+1/2)^3\over 3\omega},
\eeq
and thus the characteristic time-scale for the growth of angular speed is
\beq
t_L = \left[ {1\over \rho r^2} \sum_\ell\int d\omega {d\gamma\over d\omega}
   \exp(-\tau_0) {F_E^{(c)}(\ell+1/2)^3\over 3\omega}\right]^{-1}.
  \label{growtht}
\eeq
Figure 4 shows the growth time as a function of $r$. Note that the
growth time near the top of the radiative zone is very short (of order
a year). The growth time is inversely proportional to the energy flux in
waves, and therefore the time-scale for formation of the shear layer
will be smaller if gravity waves are generated more efficiently than
considered here (as should be the case if one considered also
the waves excited by the convective overshooting); 
the magnitude of the differential rotation within 
the layer also increases with increase in the wave flux.

The evolution of the angular speed at a given latitude below the convection 
zone is obtained by solving eq. (\ref{domega}) and the results are shown 
in fig. 5. The initial rotation speed was taken to increase with depth
below the convection zone, and the magnitude of the differential
rotation over a layer of thickness $0.01$R$_\odot$ was considered to be
$\sim 0.05$ nano-Hz initially (the mean rotation speed of the interior is 
0.4 $\mu$Hz). We show results for several different co-latitudes (60$^o$ \& 
90$^o$) and magnetic field strength (10$^3$ Gauss \& 8$\times$10$^3$ Gauss). 
The seed magnetic field we have considered lies in the horizontal plane 
along the $\hat{\bphi}$-axis, which is the expected configuration for 
fields generated by differential rotation. The initial seed toriodal 
magnetic field we start with gets amplified with time as the magnitude 
of the shear in the layer just below the convection zone increases with 
time as a result of angular momentum deposit. However, to determine the time
evolution of the magnetic field strength in a self-consistent way is
outside the scope of this work.
 
We see in fig. 5 that the time-scale for the formation of a strong shear 
layer is about 20-30 years. The time-scale for the growth of
the layer is a factor of about two smaller when the initial value of the
differential rotation across the layer is 1 nano-Hz. Moreover, the growth
time-scale is inversely proportional to the energy flux of waves.
Considering the uncertainty in the initial `seed' differential rotation
rate and the energy flux in waves, the time-scale for the formation of the 
shear layer we find should be considered to be of the same order as the solar 
cycle period.

In all of the cases shown in fig. 5, the rotation curve rises very sharply 
just below the convection zone, and this is followed by a more gently 
declining rotation curve.
Below this layer, where positive angular momentum was deposited, lies another 
layer with negative differential rotation where retrograde gravity waves are
absorbed. The sharp rise/decline of rotation in these two layers arises
as a result of the rapid increase in wave damping rate with decreasing
frequency in the local rest frame of the fluid i.e. an increase in
the rotation rate causes prograde waves to be damped over a smaller distance
scale which in turn increases the rotation rate and the feedback continues to
increase the gradient of angular velocity until much of the angular momentum
carried by prograde waves is absorbed. Since the frequencies of retrograde 
waves traveling inward in the star are increased as they propagate through
a layer with negative gradient of differential rotation, they deposit 
their negative angular momentum 
deeper in the Sun leading to the formation of another thin and strong
shear layer. The thickness of the shear layer is related to the damping
length of low frequency gravity waves which carry bulk of the angular
momentum flux. The distance from the top of the shear layer to the 
base of the convection zone is also about one damping length.

Shear layers at higher latitudes tend to be thinner and closer to the 
convection zone, and this is a result of the increase in the damping
of waves\footnote{The spectrum of waves generated by turbulent convection 
has been taken to be independent of latitude which is not strictly valid 
and could modify this dependence.}. Moreover, waves of large $m$ and $\ell$
do not propagate to high latitudes. Thus the formation of shear layer 
does not proceed at high latitudes.

For waves of a fixed frequency, an increase in the strength of magnetic 
field generally leads to increase in the damping rate and, as a result, the 
shear layer is formed closer to the convection zone, is thinner, and grows 
more rapidly (see fig. 5).

A self consistent calculation must consider the building up of the 
shear layer as well as the generation of magnetic field simultaneously. 
This is outside of the scope of this work. However, based on the physical
understanding gained for the simpler problem considered here we
expect that as the shear layer develops it twists poloidal magnetic field 
and generates toroidal magnetic field which in turn facilitates
the damping of waves and prevents the redistribution of angular momentum
in the layer by small scale turbulence (see \S3.1). Both of these effects 
speed up the growth of angular momentum gradient in the layer. Another
effect evident from figs. 5 \& 6 is that the increase in the
strength of the field causes the layer to move closer to the
convection zone. When the magnetic field becomes sufficiently strong 
it prevents the propagation of low frequency waves, which are
reflected at the top of the layer. These low frequency waves bounce back
and forth between the convection zone and the shear layer and are
absorbed after a few bounces thereby depositing their angular momentum 
and extending the shear layer upward closer to the bottom of the 
convection zone. A combination of shear instability and magnetic 
buoyancy could then lead to the ejection of the toroidal field 
from the shear layer and its deposition in the convection zone.

\medskip
\subsection{The effect of shear instability on the buildup of shear layer}

So far, we have ignored the removal of angular momentum in the shear
layer due to magnetic torque and the redistribution resulting from
hydrodynamical instabilities. 

Magnetic torque acting on the shear layer removes angular momentum
and therefore weakens the effectiveness of the gravity waves to build
up the shear. The magnetic torque per unit volume
$T_B = {\bf r}\times(\nabla\times{\bf B})\times{\bf B}/4\pi\approx
rB_r(\partial B_\phi/\partial r)/4\pi$. As long as the magnetic torque
per unit area of the shear layer, $r B_r B_\phi/4\pi$, is small
compared to the rate of angular momentum deposit in the layer
we can ignore the effect of magnetic field. Since the angular
momentum flux in prograde/retrograde  waves is of order 2x10$^{15}$ g
s$^{-2}$, magnetic torque can be neglected so long as the product
$B_r B_\phi$ is less than about 5x10$^5$ Gauss$^2$, which we assume
is the case. It should be noted that if gravity waves are excited
by some process which is more efficient than the Reynold's stress
considered in this paper, such as convective overshooting (cf. Fritts
et al. 1998), then the limit on $B_r B_\phi$ will increase correspondingly.
The limit on $B_r B_\phi$ is, however, independent of the thickness of
the shear layer since both the magnetic torque and the torque
applied by gravity waves increase as the inverse of the thickness of
the shear layer.

The shear layer is expected 
to become unstable when the gradient of rotational velocity is sufficiently
large, leading to the
redistribution of angular momentum within the layer and possible mixing 
with the overlying convection zone. 
A shear layer in a stratified medium is unstable when the so-called Richardson
criterion is satisfied (cf. Chandrasekhar 1961), 
\beq 
\frac {g}{\rho} \frac{{\rm d} \Delta\rho / {\rm d}r}{\lp {\rm d} U /
{\rm d}r\rp ^2} < Ri_c , \label{cond_instab}
\eeq
where $U$ is the horizontal velocity, $\Delta\rho$ is the density 
difference between the perturbed fluid and the ambient medium,
and $Ri_c \sim {1 \over 4}$ is the critical Richardson number 
for instability to set in.
For an adiabatic perturbation, the above condition can be written as
\beq
\frac{N^2}{\lp {\rm d} U /{\rm d}z\rp ^2} < Ri_c,
\hspace{0.5cm}
\label{Ri_dyn}
\eeq 
where $N^2 = -g(d\ln\rho/dr + g/c_s^2)$ is the Brunt-V\"ais\"al\"a 
frequency (here, we will not consider the part of the Brunt-V\"ais\"al\"a 
frequency related to the mean molecular weight stratification as it 
is very small is the Sun's outer layers).
Near the top of the radiative zone of the Sun $N/2\pi \approx 50 \mu$Hz.
Thus a shear layer of thickness $\delta r$ becomes unstable when the 
differential rotation across this layer $\delta\Omega\approx 2N \delta r/r_c$,
or when $\delta\Omega/2\pi \approx 0.1\mu$Hz for $\delta r/r_c$ of 10$^{-3}$.

Even when the Richardson criterion for instability as described above is
not satisfied, instability can still set in on small length scales due
to the weakening of the buoyancy force as a result of thermal diffusion.
The first elements to become turbulent are of size such that the time 
scale for thermal exchange with the surrounding fluid is small
and the stabilizing effect of positive entropy gradient is reduced 
(see Zahn 1992, Maeder 1995). The modified Richardson criterion which includes
heat diffusion is as follows
\beq 
 \lp \frac {\eta}{\eta +1} \rp N^2
\leq Ri_c \lp \frac{{\rm d} U }{{\rm d}r}\rp ^2 ,
\label{Rich}  
\eeq
where $\eta= v l_t/6 K$, $v$ is the velocity of turbulent elements,
$l_t$ is their size, $K$ is the thermal diffusivity defined in eq. (32), 
$\kappa$ is opacity per unit mass, and $c_p$ is specific
heat per unit mass, and $\sigma$ is the Stefan-Boltzmann constant.
The turbulent diffusivity ($D_v$) is given by those eddies which have the
largest value of $v l_t$ and which satisfy (\ref{Rich}), and it takes
the following simple form:
\beq
D_v \simeq \frac{v l_t}{3} \simeq 
2 Ri_c K \frac{({\rm d}U/{\rm d}r)^2} {N^2}
 \label{Dturb}
\eeq
(cf. Maeder 1995).
This expression remains valid 
provided $N^2/{\lp {\rm d}U/{\rm d}r\rp ^2} < Ri_c$.

We include the effect of turbulent diffusivity on the evolution of 
the shear layer and the results are shown in fig. 6. Note that the
main effect of the turbulence is to reduce the sharp velocity 
gradient at the top of the shear layer and to increase the 
thickness of the layer somewhat (compare figs. 5 \& 6)\footnote{In this
work, meridional circulation was ignored since it is known that, when
differential rotation is strong, the momentum transport is dominated
by shear turbulence (cf. Talon et al. 1997).}.

When the velocity gradient in the shear layer reaches the critical value
so that $|dU/dr|=N Ri_c^{1/2}$, the nature of the instability changes. 
The thermal diffusion is no longer required to destabilize
the fluid and a dynamical instability sets in whose growth rate is 
of order of the rotation period (see e.g. Mac Donald 1983).
The shear layer, which lies just below the convection zone, is likely to
merge with the convection zone depositing the magnetic field that was 
build up in the layer as a result of differential rotation. This could 
be the start of a new magnetic cycle.
That situation was never reached in our calculations since the diffusive
shear instability was always large enough to redistribute angular momentum
before condition (\ref{Ri_dyn}) was satisfied.
However, due to the lack of a reliable model, 
we have ignored the stabilizing effect of magnetic fields on the
shear instability. We expect that magnetic fields will reduce the
value for the turbulent diffusion coefficients we have used here,
and therefore radial differential rotation will build up to larger 
values than shown in fig. 6 resulting in a more efficient generation 
of a toroidal magnetic field. In this case the dynamical instability 
condition is likely to be satisfied. The shear layer is also more likely
to become dynamically unstable if the angular momentum flux in gravity
waves is larger than estimated in \S2.3.

The time-scale for the buildup of the shear layer and its distance from the
base of the convection zone decreases with increasing latitude. Thus, dynamical
instability and mixing of layer is expected to first occur at high latitudes
and later in the equatorial region. 
However, a complete description of the latitude dependence would require
a truly two dimensional model. Indeed, as there is no stabilizing stratification
along horizontal layers, horizontal shear instabilities are much stronger
than their vertical analogs, and we expect important horizontal redistribution
of angular momentum to occur.

\medskip
\section{Discussion}

The gravity waves generated by turbulent convection
are potentially very efficient in redistributing angular momentum
in the radiative interior of the Sun. Most of the angular momentum
flux is carried by the lowest frequency waves, which have periods
of the order of the convective time-scale at the bottom of the convection 
zone (or $\sim 0.3 \mu$Hz for the present day Sun). The total energy 
flux in these low period gravity waves is of the order of 0.03\% of the
solar luminosity and the angular momentum flux ($F_L$) in prograde waves 
is $\sim 2 \times 10^{15}$ g s$^{-2}$.
We assume that prograde waves, which carry positive angular momentum flux,
and retrograde gravity waves, which have negative angular momentum flux,
are excited to the same amplitude and thus, the total net angular momentum
flux at the top of the radiative interior is zero; in this case, gravity 
waves alone merely redistribute angular momentum in the interior.

Kumar \& Quataert (1997) and Zahn, Talon \& Matias (1997)
suggested that gravity waves generated by turbulent convection
could bring the radiative interior of the Sun in corotation with
the convection zone. The work presented here corrects a sign error,
and adds one crucial element to their picture that was left out i.e. the
formation of a double shear layer below the convection zone
filters out either prograde or retrograde waves depending
on the sign of the initial velocity gradient below the convection zone.

The thickness of these shear layers is of the order of the 
wave damping length (which is a few percent of the pressure scale
height). The magnitude of the differential rotation ($|\delta\Omega|/2\pi$)
in the shear layers reaches about 0.4 $\mu$Hz when turbulent diffusion of
angular momentum is not included. The magnitude of $\delta\Omega/2\pi$ can be
estimated using the relation $|\delta\Omega|\sim t F_L/(\rho r^2\delta r)$
(this is applicable when waves with one sign of angular
momentum are absorbed in the layer), and we find it to be 0.1 $\mu$Hz after
a period of a few years when the thickness of the shear layer is
$3 \times 10^{-3} R_\odot$. The thickness is uncertain by a factor
of a few due to the uncertainty in spectrum of gravity waves and 
the angular momentum redistribution as a result of shear and magnetic
instability.

As the angular velocity increases in the layer, it becomes unstable
to small length scale perturbations which cause mixing of the layer, 
and thus contributes to the mixing of elements near the top of the
radiative interior of the Sun as seems to be required by the
helioseismic inversion of the sound speed (Basu, 1997).

If only ``diffusive'' turbulence is present (as in our calculations),
the first shear layer lying at the top of the radiative zone will 
eventually merge with the surface convection
zone and the second shear layer, of opposite sign, will replace it
at the top of the radiative zone, reversing the structure, much as
in the biennial oscillation, which is well known in atmospheric sciences
(cf. Holton \& Lindzen, 1972).
Taking into account the stabilizing effect of a toroidal magnetic field
could change this situation as stronger shears would result, leading
perhaps to the appearance of a more drastic ``dynamical'' instability
which might destroy much of the double shear layer.

Most of the angular momentum deposited in these shear layers is transferred 
back into the convection zone once they become turbulent, but further work 
is needed to check whether there will be a net amount of angular
momentum taken out of the radiation zone, 
and how it will depend on the angular velocity gradient. This issue must 
also be settled in order to get a clear picture of the interaction 
between these short-lived shear layers and the quasi-stationary tachocline, 
which is the somewhat thicker transition layer that links the 
differential rotation of the convection zone to the almost uniform 
rotation below (see Spiegel \& Zahn 1992).

A poloidal magnetic field threading the shear layer will get twisted 
as the differential
rotation in the layer increases and, as a result, toroidal magnetic
field will be generated. The increasing toroidal field in turn 
accelerates the rate of angular momentum deposit in the layer, and 
inhibits small length scale shear instability which reduces the rate
of growth of the angular momentum gradient. Once the magnetic
field in the shear layer becomes strong enough to reflect low
frequency or high azimuthal wavenumber waves, these waves travel
back and forth between the shear layer and the convection zone
and their angular momentum is absorbed in this intermediate layer
in a few transit time. The result of this is to extend the shear layer
closer to the convection zone. At some stage in its evolution,
the shear layer becomes unstable and mixes with the convection zone 
and the magnetic field is the layer floats upward to start a new
magnetic (half) cycle. Once the layer has relaxed and settled back to
a stable configuration, after having given up much of the angular
momentum and magnetic field that it had contained, the whole process 
starts all over again. Let us point out that this model predicts cyclic 
reversals of the toroidal field with a steady poloidal field at the base 
of the convection zone, which could well be a fossil field, whereas in 
classical dynamo theory the poloidal field too undergoes such reversals.
We note also that waves of large azimuthal order ($m$) do not reach 
high latitude. Therefore, the formation of the shear layer and 
the generation of magnetic field in not likely to occur at high
latitudes which is consistent with the lack of sunspots there.

Incorporating these various processes in a self consistent manner in
the evolution calculation is, however, very complicated and has not 
been considered here. But it is tempting to link the cyclic build-up of 
these shear layers, which produce toroidal field of alternating sign, 
with the solar magnetic cycle, since the time-scales are of the same order. 
Future helioseismic observations and inversions with improved resolution 
should be able to determine the presence of a double shear layer at the 
top of the radiative interior predicted here.

{\bf Acknowledgment: } PK thanks Peter Gilman for helpful discussion; 
ST and JPZ are grateful to Olivier Ringot for helping them to clarify
the issue of angular momentum transport by gravity waves. We thank
Evry Schatzman and an anonymous referee for a careful reading of
the manuscript and for suggestions for improvements.

\vfill\eject

\vfill\eject
\bigskip
\centerline{\bf Figure Captions}
\bigskip

\noindent FIG. 1.--- The radial wavenumber ($k_r R_\odot$) as a 
function of wave frequency for $B=10^3$, \& 8x$10^3$ Gauss, and for 
latitude ($l_\theta$)= 0$^o$ \& 30$^o$. The solid line is for 
($B=10^3, l_\theta=0^o$), the dotted line ($B=10^3, l_\theta=30^o$),
dashed line for ($B=8\times 10^3, l_\theta=0^o$), and the dash-dot
line is for ($B=8\times 10^3, l_\theta=30^o$). In all of these
cases the horizontal wave number was taken to be $k_h=50/R_\odot$,
$m=30$, $\Omega/2\pi=0.4\mu$Hz, $\rho=0.2$ g cm$^{-3}$, and the
Brunt-V\"ais\"al\"a frequency $N=100\mu$Hz (the last three values correspond
to the top of the solar radiative zone).

\medskip
\noindent FIG 2.--- The left panel shows energy luminosity per unit
radian frequency in gravity waves of degree $\ell=1$ as a function of 
frequency (energy luminosity is defined to be the product of
energy flux and the surface area). The right panel shows the
frequency integrated wave action, or angular momentum flux divided by $m$,
as a function of $\ell$; the frequency integration was carried out over
the interval 0.6--3.0 $\mu$Hz (most of the contribution to the wave action
comes from frequency close to 0.6 $\mu$Hz, see the left panel).
We have used the solar model of J. Christensen-Dalsgaard in this
as well as all the other numerical calculations shown in this paper.

\medskip
\noindent FIG 3.--- Damping length for gravity waves divided by the 
wavelength of the wave is shown as a function of wave frequency for 
$\ell=2$ (solid line), 15 (dotted line), and 50 (dashed line).

\medskip
\noindent FIG 4.--- The time-scale for exponential growth of angular velocity
gradient below the solar convection zone as a function of $r$ for two different
values of the magnetic field $B=10^3$, \& 8x$10^3$ Gauss, and 
latitude ($l_\theta=$) 0$^o$ \& 30$^o$. The top of the radiative zone is at
$r/R_\odot=0.71195$. The growth time-scale was calculated by integrating
over frequency range of 0.6--3.0 $\mu$Hz and the maximum $\ell$ was taken to be
55. The solid line is for
($B=10^3, l_\theta=0^o$), the dotted line ($B=10^3, l_\theta=30^o$),
dashed line for ($B=8\times 10^3, l_\theta=0^o$), and the dash-dot
line is for ($B=8\times 10^3, l_\theta=30^o$).

\medskip
\noindent FIG 5.--- Evolution of the shear layer at two different
latitudes (0$^o$ and 30$^o$) and for $B=10^3$ and 8x10$^3$
Gauss. The initial rotation profile is taken to increase with depth 
and the magnitude of the initial differential rotation is about 2 nano-Hz
over the range of $r$ shown in these plots. The thin solid line
is the rotation profile after 10 yr, dotted line at 20 yr, dashed line
at 30 yr and dash-dot line at 40yr. 
Turbulent diffusion of angular
momentum in the shear layer is not included in these calculations.

\medskip
\noindent FIG 6.--- Evolution of the shear layer.
The top left pannel shows the difference between
a calculation including turbulent diffusion (thin continuous line)
and excluding turbulent diffusion (thick dashed line) at the
equator and for $B=10^3$ Gauss at 20, 30 and
40 years. The other three pannel show the shear layer including
turbulent diffusion at 0$^o$ and 30$^o$
and $B=10^3$ and 8x10$^3$ Gauss. The initial rotation profile is as
in Fig. 5. Ages are indicated on the plot.

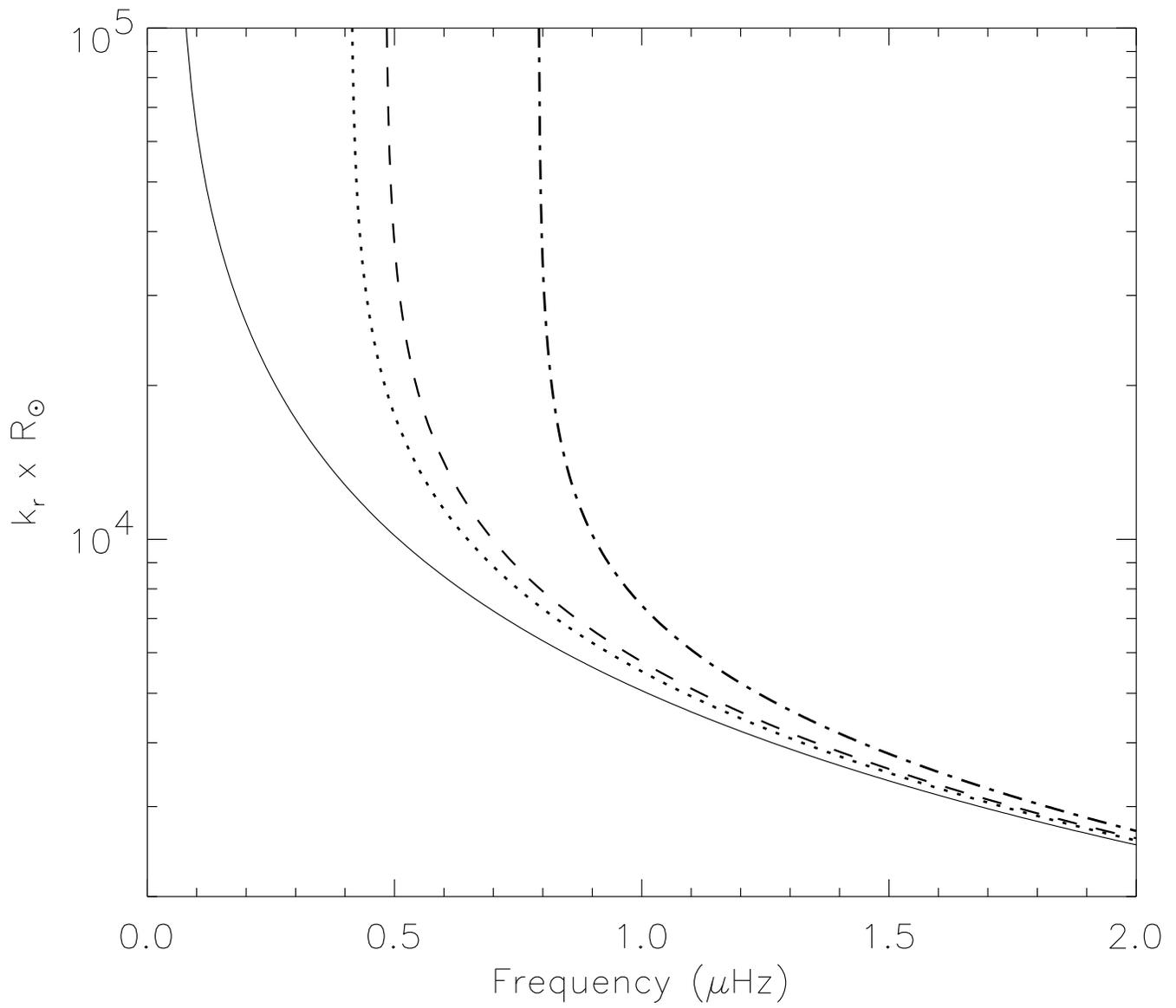
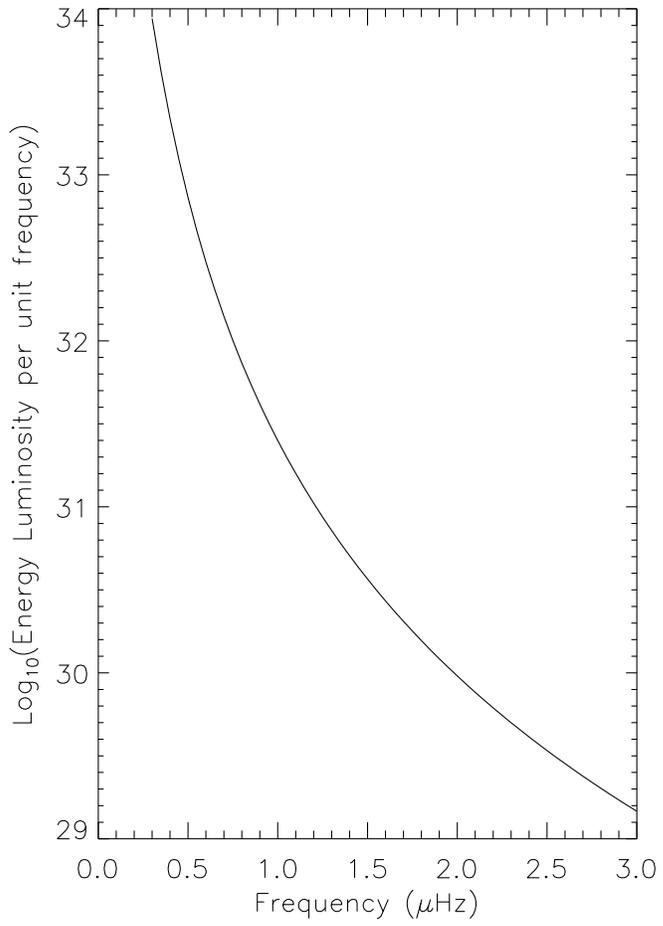
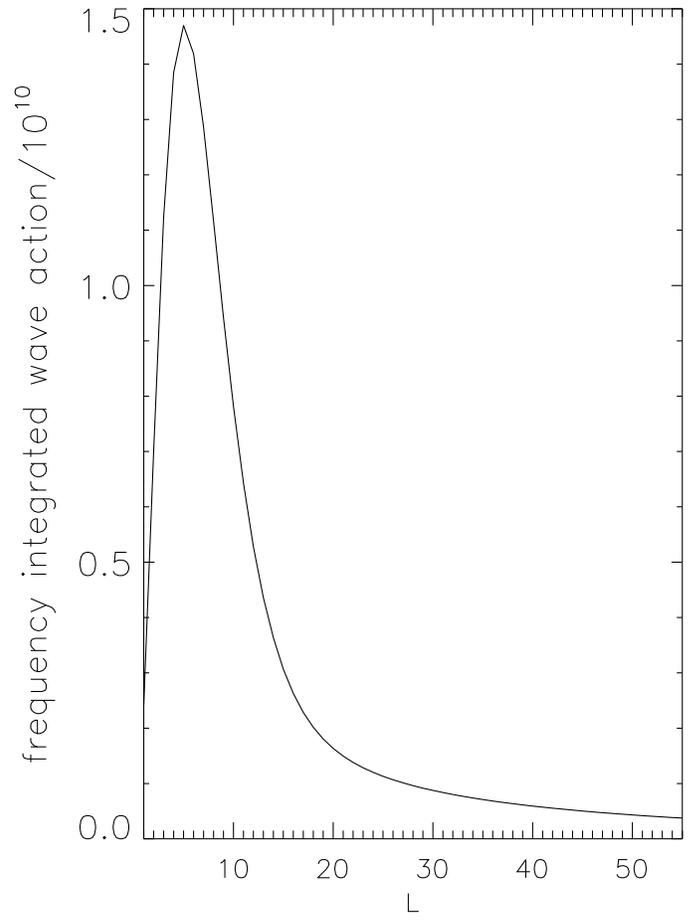
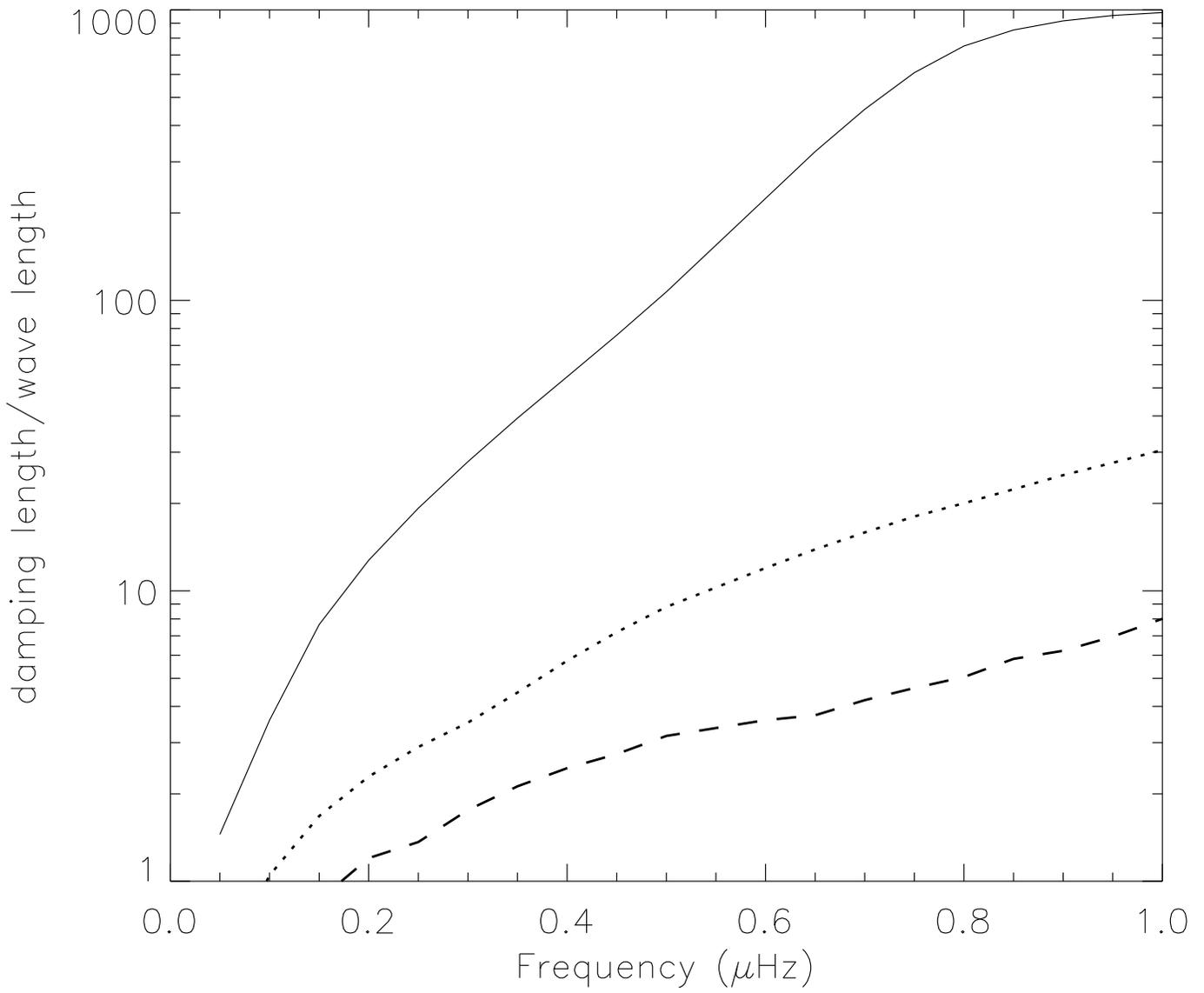
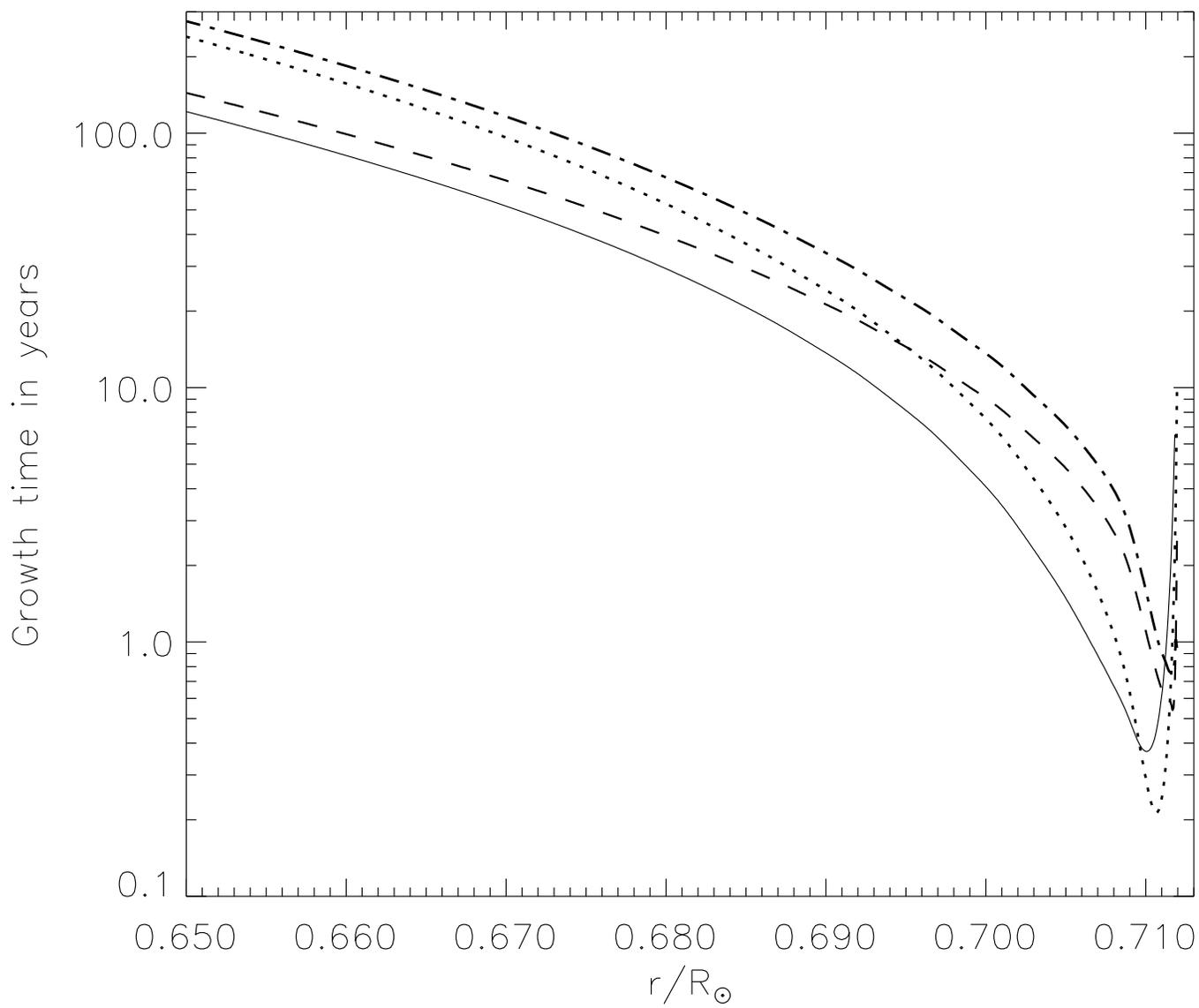
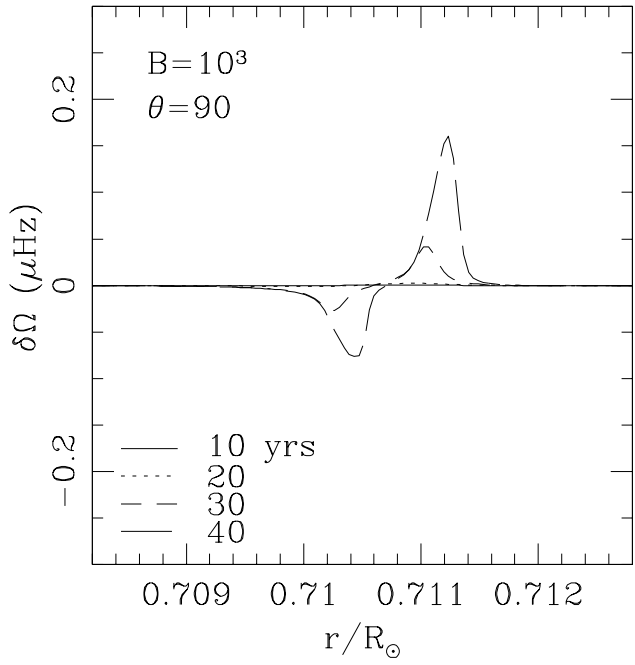
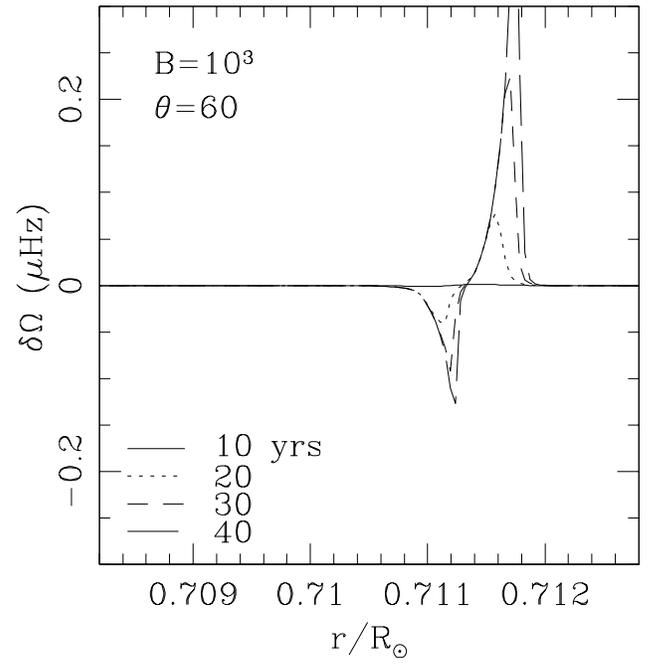
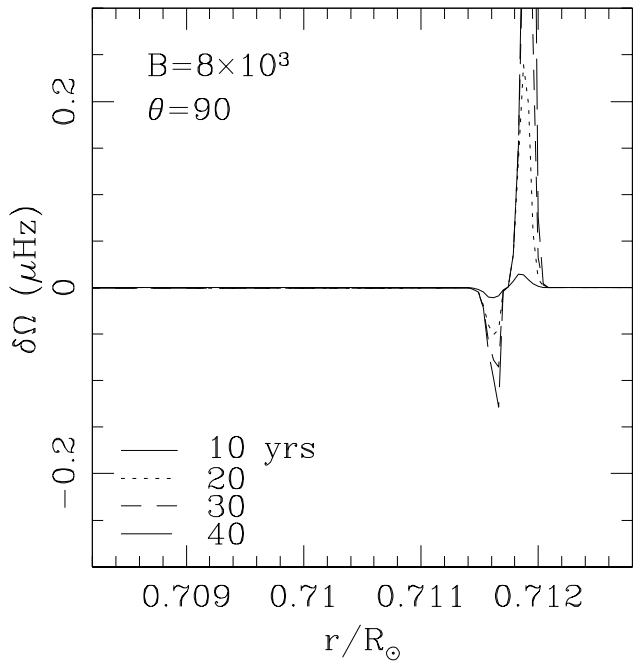
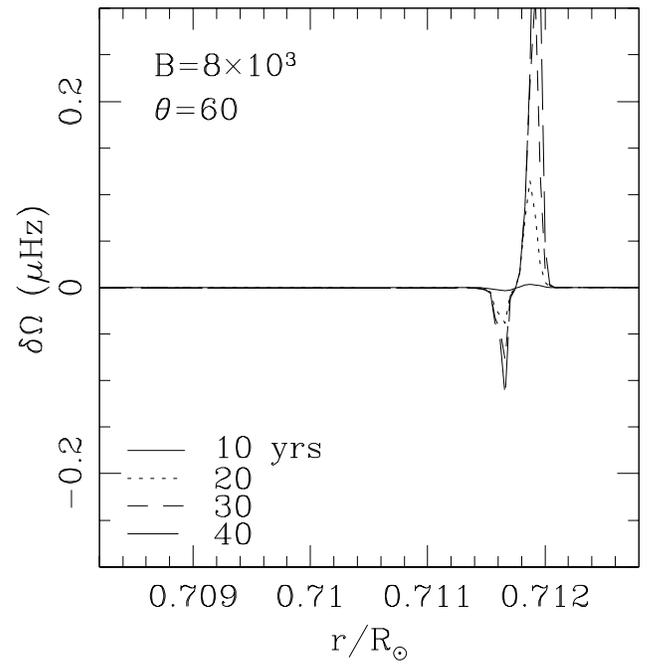
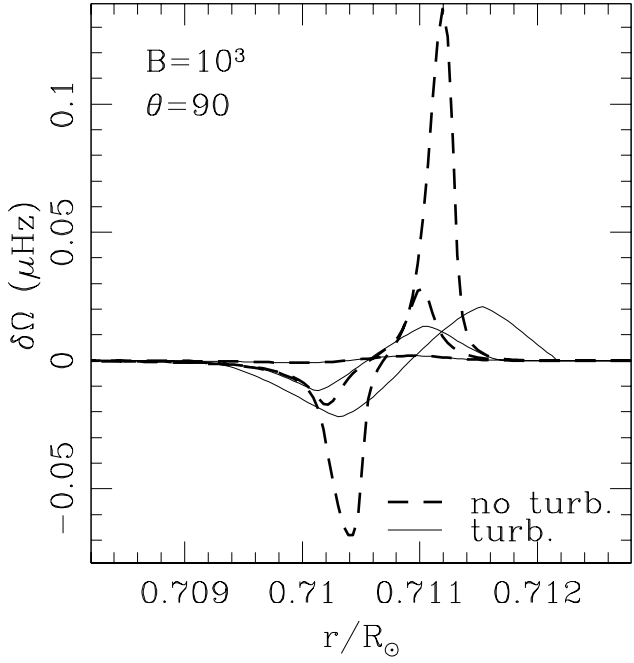
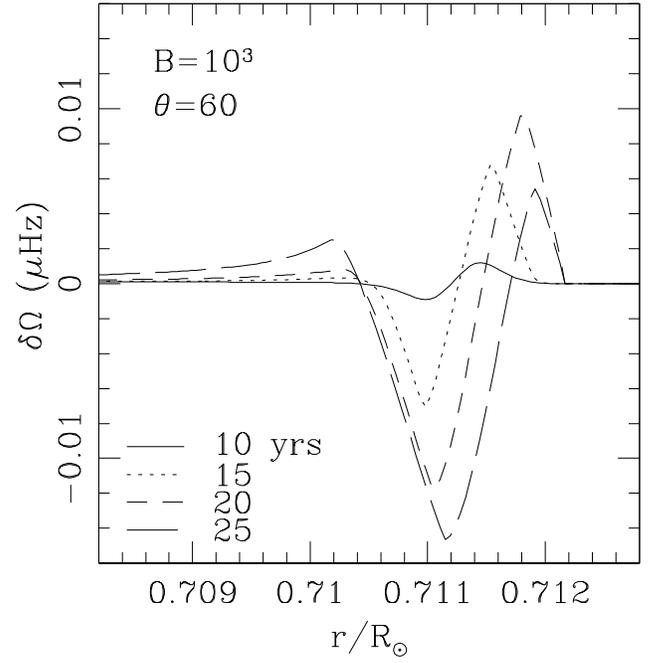
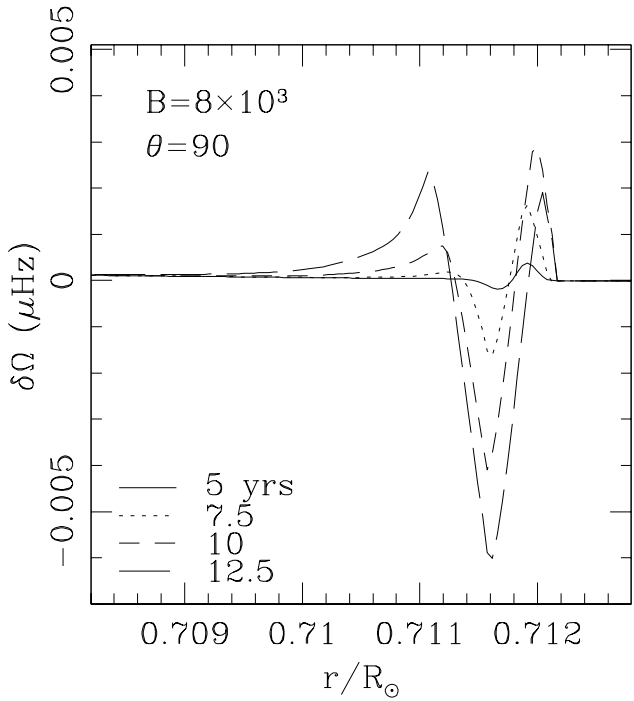
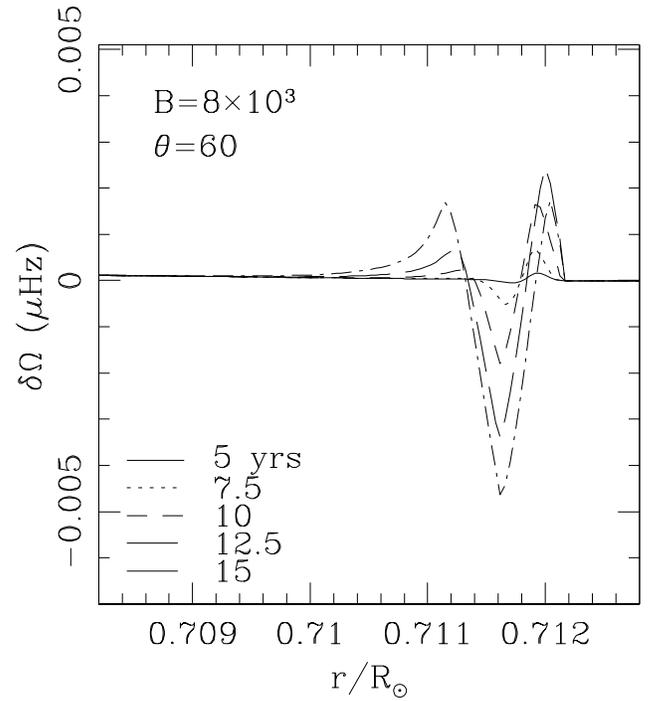
\end{document}